\documentclass[aps,prl,twocolumn,10pt,nofootinbib,superscriptaddress,preprintnumbers]{revtex4-2}

\usepackage[utf8]{inputenc}
\usepackage[T1]{fontenc}
\usepackage{amssymb,amsfonts,amsmath}
\usepackage{graphicx}
\usepackage{hyperref}
\usepackage{nicefrac}
\usepackage{xcolor}
\usepackage{booktabs}
\usepackage{float}
\usepackage{placeins}

\begin{document}
\preprint{MITP-25-035}

\title{Probing how bright the quark-gluon plasma glows in lattice QCD}

\author{Ardit Krasniqi$^{*}$}
 \affiliation{PRISMA$^+$ Cluster of Excellence  \& Institut f\"ur Kernphysik,
Johannes Gutenberg-Universit\"at Mainz, D-55099 Mainz, Germany}

\author{Marco C\`e}
\affiliation{Dipartimento di Fisica, Universit\`a di Milano-Bicocca and
INFN, Sezione di Milano-Bicocca,
Piazza della Scienza 3, 20126 Milano, Italy}

\author{Tim Harris}
\affiliation{Institute for Theoretical Physics, ETH Z\"urich,
Wolfgang-Pauli-Str.~27, 8093 Z\"urich, Switzerland}

\author{Renwick J.~Hudspith}
\affiliation{Department of Physics, Carnegie Mellon University, Pittsburgh,
Pennsylvania 15213, USA}

\author{Harvey B.~Meyer}
 \affiliation{PRISMA$^+$ Cluster of Excellence  \& Institut f\"ur Kernphysik,
Johannes Gutenberg-Universit\"at Mainz, D-55099 Mainz, Germany}
\affiliation{Helmholtz-Institut Mainz, Johannes Gutenberg-Universit\"at Mainz,
D-55099 Mainz, Germany}

\date{May 2025}

\begin{abstract}

\noindent Determining the spectrum of photons emitted by the quark-gluon plasma non-perturbatively remains an open computational challenge.
In this letter we calculate two moments of that spectrum at a tem\-pe\-rature $T\approx$ 254\,MeV, employing lattice QCD with two flavors of $\mathcal{O}(a)$-improved Wilson fermions, without facing an inverse problem.
Our central value for the difference of these two moments, which is sensitive to photon energies $\omega\gtrsim \pi T$, is lower than, but compatible with that obtained by integrating the leading-order weak-coupling photon spectrum.
This study informs the \textit{direct photon puzzle} in heavy-ion collision phenomenology, where it has proved difficult to simultaneously explain the 
yield and azimuthal anisotropy of photons not originating from final-state hadronic decays.

\end{abstract}
\maketitle

\textit{Introduction.---}The radiation spontaneously emitted  by a thermal medium provides important insight into its composition and dynamics. 
For instance, in astronomy the spectral lines of interstellar gas are the main source of information on its constituents and overall motion~\cite{Draine:2010}.
In this letter, we are concerned with the photon emissivity of strongly-interacting matter above the (crossover) transition temperature $T_\mathrm{c}$. At asymptotically-high temperatures,
from a plasma of quasi-free moving quarks and gluons, a photon can only be emitted if a quark enters a collision, making the photon emissivity directly sensitive to interactions within the medium. 

In relativistic heavy-ion collisions (HIC), the transverse momentum dependent yield and azimuthal anisotropy of direct photons\footnote{Direct photons are those that are not produced through the decay of final-state hadrons, as for instance in $\pi^0\to\gamma\gamma$.} have been studied in several experiments~\cite{PHENIX:2008uif,PHENIX:2011oxq,PHENIX:2012jbv,PHENIX:2014nkk,PHENIX:2015igl,PHENIX:2022rsx,PHENIX:2025ejr,STAR:2016use,ALICE:2015xmh,ALICE:2018dti}. These observables are sensitive to photon production throughout the spacetime history of the matter produced in the collision, and predicting them requires a complex modeling  of the system's expansion and cooling.
It has been difficult to explain the experimental measurements by the (otherwise very successful) hydrodynamic description and a model of the thermal photon emissivity~\cite{Gale:2021emg} based on the leading weak-coupling prediction for the high-temperature phase. This open problem has been dubbed the ``direct photon puzzle''~\cite{David:2019wpt,Chatterjee:2024blo,Gale:2025ome}.

The differential photon emissivity of a medium at perfect thermal equilibrium is, to leading order in the fine-structure constant $\alpha$, proportional to a thermal spectral function $\sigma_{\rm em}(\omega)$, $\omega$ being the photon energy in the rest frame of the fluid.
In this letter, we compute two energy moments of this  spectral function at a temperature $T\approx 254\,$MeV, i.e.~around $1.2\,T_\mathrm{c}$~\cite{Brandt:2016daq} from first principles using lattice QCD. 
We compare our results with the moments obtained by integrating the full
leading-order weak-coupling spectral function derived by Arnold, Moore and Yaffe
(AMY) from kinetic theory~\cite{Arnold:2001ms}.
Our work tests the adequacy of the latter at temperatures typical of heavy-ion collisions. Thereby, it contributes to our theoretical understanding of thermal strong-interaction matter as well as to resolving the direct photon puzzle~\cite{David:2019wpt} of HIC phenomenology.

In contrast to previous lattice QCD studies of the photon emissivity~\cite{Ghiglieri:2016tvj,Ce:2020tmx,Ce:2022fot,Ali:2024xae}, the present study does not involve tackling a numerically ill-posed inverse problem. It significantly improves upon our recent works~\cite{Torok:2022vki,Ce:2023oak} through superior statistical precision on the current-current correlation functions, thereby allowing us  to more stringently probe the emission of the phenomenologically most relevant photons, $\omega\gtrsim \pi T$.

\textit{Framework.---}The differential rate of photon emission per unit volume plasma at temperature $T=1/\beta$ can be expressed to leading order in $\alpha$ as~\cite{McLerran:1984ay}
\begin{align}
    \frac{\mathrm{d}\Gamma}{\mathrm{d}\omega} = \frac{\alpha}{\pi} \, 
    \frac{2\omega\,\sigma_{\rm em}(\omega)}{e^{\beta\omega}-1}\;.
\end{align}
The photon spectral function $\sigma_{\rm em}(\omega)$  is rich in information about the thermal system. 
In particular, the electric conductivity is given by 
$\sigma_{\rm el} 
= 2\pi\alpha \lim_{\omega\to0} \sigma_{\rm em}(\omega)/\omega$.
In the Matsubara formalism of thermal field theory, in which the (imaginary) time variable is compact, $0\leq x_0 < \beta$, we define for each frequency $\omega_n = 2n\pi T$, $n\in\mathbb{Z}$, the correlation function of the vector current $j_\mu(x)$,
\begin{align}
    H_E(\omega_n) = \int \mathrm{d}^4x\; e^{\omega_n(ix_0 - x_1)}\;
    \left\langle j_3(x) j_3^\dagger(0)\right\rangle\;. 
\end{align}
The quantity $H_E(\omega_n)$ has a spectral representation at fixed, lightlike virtuality \cite{Meyer:2018xpt},
\begin{align}
    H_E(\omega_n) = -\frac{\omega_n^2}{\pi}\int_0^\infty \frac{\mathrm{d}\omega}{\omega}\, \frac{\sigma(\omega)}{\omega^2+\omega_n^2}\;.
\end{align}
The difficulty of computing $H_E(\omega_n)$ increases dramatically with $n$ due to the enhanced contribution at large spatial separations $x_1$, where the correlator suffers an exponential deterioration of the signal-to-noise ratio.
The first moment $H_E(\omega_1)$ was determined with an uncertainty of one percent in Ref.~\cite{Ce:2023oak}.
Of particular interest is the difference of the $n=2$ and $n=1$ moments, since it suppresses the contribution of the very soft photons and is sensitive to the kinematically interesting region $\omega\gtrsim\pi T\approx 1\,\mathrm{GeV}$ relevant to the direct photon puzzle.
Therefore, we focus on the computation of the $n=2$ moment in the following.

We use simulations with $N_\mathrm{f}=2$ flavors of dynamical $\mathcal{O}(a)$-improved Wilson quarks with the Wilson gauge action (see End Matter), and

compute the correlator of the isospin current\footnote{
For $j_\mu:=\frac{2}{3}\bar u \gamma_\mu u - \frac{1}{3}\bar d \gamma_\mu d + \dots$, 
we have $\sigma_{\rm em}=\sigma$.
With $j_\mu$ set to the isospin current, neglecting SU(3)-flavor breaking effects in the high-temperature phase, we have $\sigma_{\rm em}\simeq (\frac{4}{9}+\frac{1}{9}+\frac{1}{9})\sigma$.} $j_\mu:=(\bar u \gamma_\mu u - \bar d \gamma_\mu d)/\sqrt{2}$.
The specific expression used in the lattice calculation is~\cite{Ce:2023oak}
\begin{align}
\label{eq:H_E_standard_subtraction}
&  H_E(\omega_n) = a^4 \sum_{x}
  \left( e^{i\omega_n x_0} - e^{i\omega_n x_2} \right) 
  e^{-\Omega_n x_1}
  \left\langle j_3(x) j_3^\dagger(0)\right\rangle \nonumber 
  \\  &  \qquad ~\;\quad = \frac{a}{2} \sum_{x_1=-\infty}^\infty h_E(\omega_n,x_1) \,,
\\ & h_E(\omega_n,x_1) = 2\left[G_{\text{ns}}^T(\omega_n,x_1)-G_{\text{st}}^T(\omega_n,x_1)\right]\cosh(\Omega_n x_1)\,,
\nonumber
\end{align}
where $G_{\text{ns}}^T$ and $G_{\text{st}}^T$ denote non-static and static transverse screening correlators with momentum insertions along the temporal and spatial directions, respectively. In the continuum, the static contribution vanishes, but on the lattice it helps reduce discretization errors. Also, one may choose $\Omega_n=\omega_n(1+{\mathcal{O}}(a^2))$ in order to suppress artefacts due to the finite lattice spacing $a$~\cite{Meyer:2021jjr}.

\begin{figure}[t]
        \includegraphics[scale=0.53]{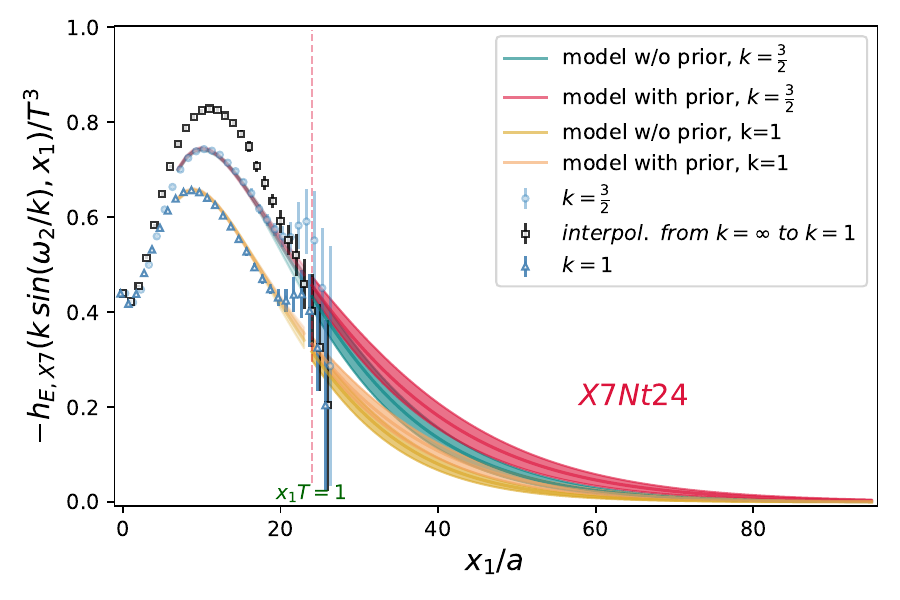}
	\includegraphics[scale=0.33]
    {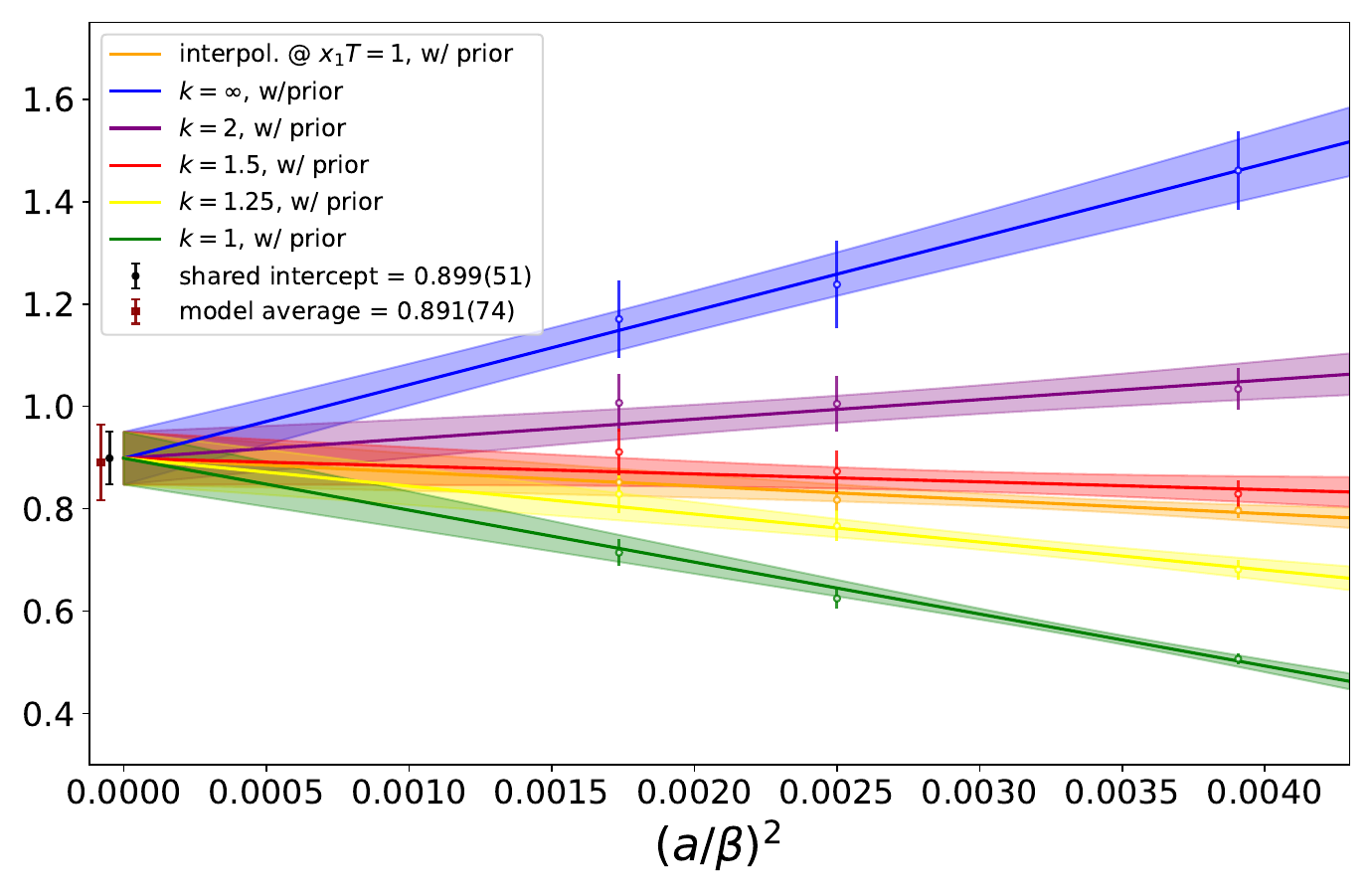}
        \caption{\textbf{Top panel:} Integrands $-h_E(\Omega_n(k),x_1)/T^3$ for $k\in \{1,\frac{3}{2},\infty\}$ along with the interpolation between the prescriptions $k=\infty$ and $k=1$ at $x_1T=1$ and choosing $d\approx0.15\,$fm in Eq.~\eqref{eq:smooth_heaviside}. The integrands are shown on the X7 ensemble together with the tail modification.
      \textbf{Bottom panel:} Simultaneous (uncorrelated) continuum extrapolation with a shared intercept for $-H_E(\omega_2)/T^2$ using priors from the ``ratio method''. The result $[-H_E(\omega_2)/T^2]_{\text{w/ prior}}^{\text{uncorr.}} = 0.899(51)_{\text{stat}}\,$ agrees well with our quoted MAV result, Eq.\,(\ref{eq:H_E_omega2_result_with_prior_mav}).}
    \label{fig:H_E_omega_2_with_tails_d3_X7_and_simul_cont_extr_H_E_omega_2}
\end{figure}

A one-parameter family of kernel regularizations is given by the choice
\begin{equation}
    \label{eq:sine_modification}
    \Omega_n(k) =\frac{k}{a}\text{sin}\left(\frac{a\omega_n}{k}\right)\,,\quad k\in \mathbb{R}.
\end{equation}
In our final analysis, we include continuum extrapolations for $k = 2$, a choice motivated by the dispersion relation of a free lattice scalar field,
 and $k = 3/2$, which empirically yields the flattest continuum extrapolation (see Fig.~\ref{fig:H_E_omega_2_with_tails_d3_X7_and_simul_cont_extr_H_E_omega_2}, bottom panel). 
 As a third variant for $\Omega_n$, 
 we opt for a smooth interpolation centered at $x_1=\beta$ between the values $\omega_n$ at short distances and $\frac{1}{a}\sin(a\omega_n)$ at long distances.
 This choice is guided by a study of cutoff effects in leading-order lattice perturbation theory; see End Matter. 
 
Our lattice simulations are performed on three ensembles, labeled X7, W7, and O7, with lattice spacings $a=0.033, 0.039, 0.049\,\mathrm{fm}$, tuned to a common temperature of $T \approx 254\,\mathrm{MeV}$.
We achieve very high statistics on the relevant correlators: the propagators
entering Eq.~\eqref{eq:H_E_standard_subtraction} are computed using  ``Z2SEMWall''~\cite{Boyle:2008rh,ETM:2007xow,McNeile:2006bz} stochastic momentum~\cite{Gockeler:1998ye} wall sources, within a truncated solver approach~\cite{Bali:2009hu}.
That is, we perform many more low-precision solves for translations in space across the lattice than high-precision ones. 
The DFL+SAP+GCR algorithm of~\cite{Luscher:2007se} is used to solve the Dirac equation, whereby the cost of generating the deflation subspace is amortized by the large number of low-precision translations and stochastic ``hits'' we perform.
The combination of stochastic momentum wall sources and the truncated solver method allows us to achieve much higher statistical resolution than the exact point sources previously employed in~\cite{Ce:2023oak} with typically lower overall cost.
The only limitation is that the momenta are fixed \emph{a priori}.

We perform simultaneous, correlated continuum extrapolations for two out of the three data sets associated with a particular choice of $\Omega_n$, enforcing a common continuum limit between them. For each of the three pairs, we consider four combinations depending on whether the coarsest lattice spacing ensemble labelled O7 is included or excluded independently in each extrapolation (i.e., both extrapolations include O7, only one includes O7, or both exclude it). This results in $3 \times 4 = 12$ fit models. 

We average the results derived from these different choices according to the Akaike Information
Criterion~\cite{Akaike:1998zah} and the model averaging method from Refs.~\cite{Jay:2020jkz, Kuberski:2024bcj}.
The weight assigned to each fit is
\begin{equation}
w_i = N \exp\left[ -\frac{1}{2} (\chi_i^2 + 2k_i - 2n_i )\right],
\label{eq:weights}
\end{equation}
where \( k_i \) is the number of fit parameters and \( n_i \) is the number of data points in the fit with minimized \( \chi_i^2 \). The normalization \( N \) is  such that \( \sum_i w_i = 1 \).
The central value and statistical uncertainty of an observable $\mathcal{O}$ are obtained via a weighted average over all models:
\begin{equation}
\bar{\mathcal{O}} = \sum_i w_i \mathcal{O}_i\,,
\label{eq:central_value}
\end{equation}
while the systematic error associated with the continuum extrapolation is estimated as
\begin{equation}
(\delta \mathcal{O})_{\text{sys}}^2 = \sum_i w_i (\mathcal{O}_i - \bar{\mathcal{O}})^2.
\label{eq:systematic_error}
\end{equation}

\begin{figure}[t]
    \includegraphics[scale=0.53]{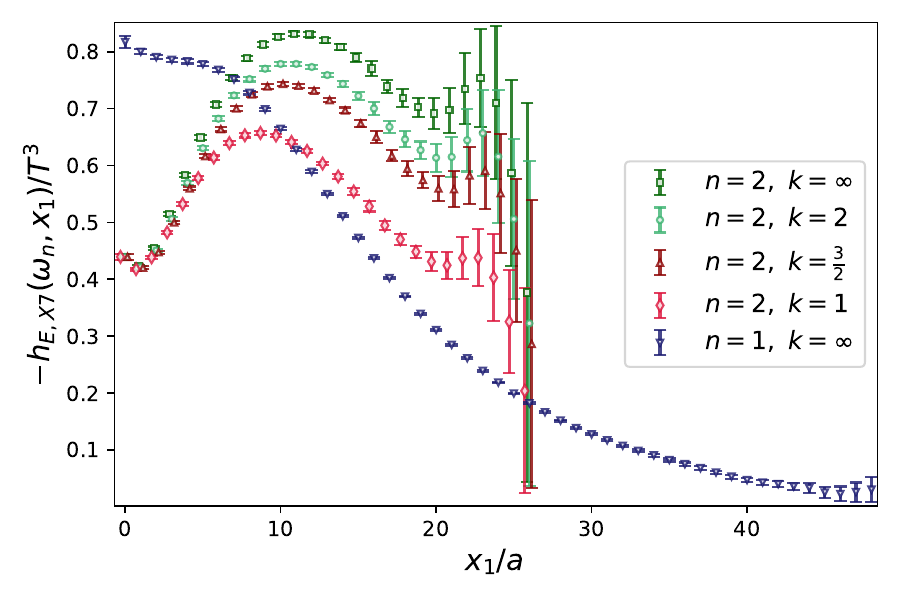}
        \caption{Integrands $-h_E(\omega_2,x_1)$ with different kernel modifications based on the parameter $k$ as well as $-h_E(\omega_1,x_1)$ on the finest ensemble X7.}
    \label{fig:H_E_omega_2_different_k_X7}
\end{figure}
The statistical uncertainties of our (derived) observables are estimated using the $\Gamma$ method~\cite{Madras:1988ei,Wolff:2003sm,Ramos:2018vgu}, as implemented in the \texttt{pyerrors} package~\cite{Joswig:2022qfe}.

\textit{Results.---}The numerical challenge to determine the moments $H_E(\omega_n)$ stems from (a) the exponentially deteriorating signal-to-noise ratio for the correlators at large distances $x_1$ and (b) the exponential enhancement of the tail by the kernel.
 As a result, the uncertainty of the integrand grows exponentially at large source-sink separations; see Fig.~\ref{fig:H_E_omega_2_different_k_X7}. 
 However, it is known that both correlators $G_{i}^T$,  $i\in\{{\rm ns},\,{\rm st}\}$, are given at long distance (and in infinite volume) by a sum of exponentials with positive prefactors. 
 Taking into account the periodic boundary conditions in our simulations, we employ the fit ansatz 
\begin{align}
\label{eq:inf_volume_fit}
    G_i^T(\omega_n,x_1) = \sum_{l=0}^2 \vert A_i^l \vert^2\, \text{cosh}[(x_1-L/2)\,m^l_i(\omega_n)]\,,
\end{align}
The pairs $m^l_i,\,A_i^l$ represent the $l$th screening masses and
corresponding amplitudes in a given sector $\omega_n$. 
The impact of including additional states in the fit ansatz~\eqref{eq:inf_volume_fit} is illustrated in Fig.~\ref{fig:fit_systematic} (in End Matter), where results from one-, two-, and three-state fits are compared.

\begin{figure}[t]
    \includegraphics[scale=0.53]{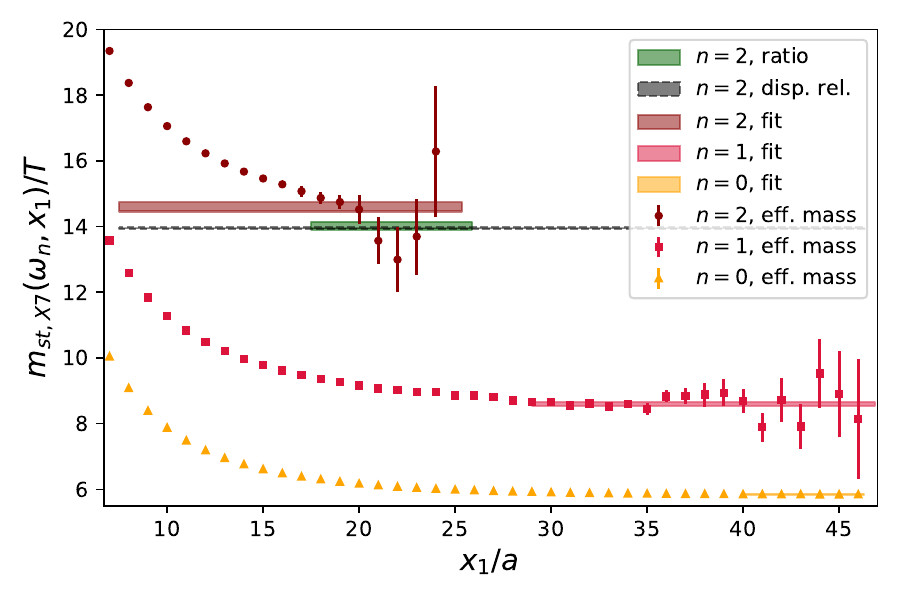}
    \includegraphics[scale=0.53]{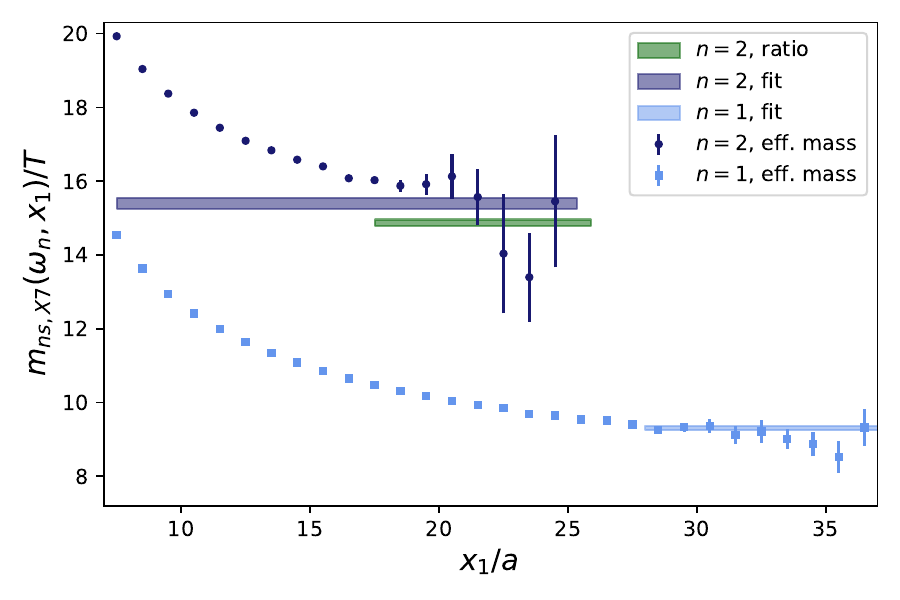}
    \caption{\textbf{Top panel:} Effective mass (see Eq.~\eqref{eq:effmass}) and fit results for the static correlators at spatial momenta $\omega_0,\omega_1$ and $\omega_2$ on the X7 ensemble. \textbf{Bottom panel:}  Effective mass and fit results for the non-static screening correlator in the $n=1$ and $n=2$ Matsubara sector.}
    \label{fig:static_and_non_static_eff_masses_X7}
\end{figure}

The results for the ground states in the $n=2$ sector, after taking the continuum limit are
\begin{align}  
    \label{eq:screening_mass_static_omega2_from_fit}  
     m_{\text{st}}^{\text{0,fit}}(\omega_2)/T &= 14.73(26)\,,  
     \\
     \label{eq:screening_mass_nonstatic_omega2_from_fit} 
     m_{\text{ns}}^{\text{0,fit}}(\omega_2)/T &= 15.72(31)\,.
\end{align}  
The corresponding results for our finest ensemble are  displayed in Fig.~\ref{fig:static_and_non_static_eff_masses_X7}.

\begin{figure}[t]
    \includegraphics[scale=0.53]{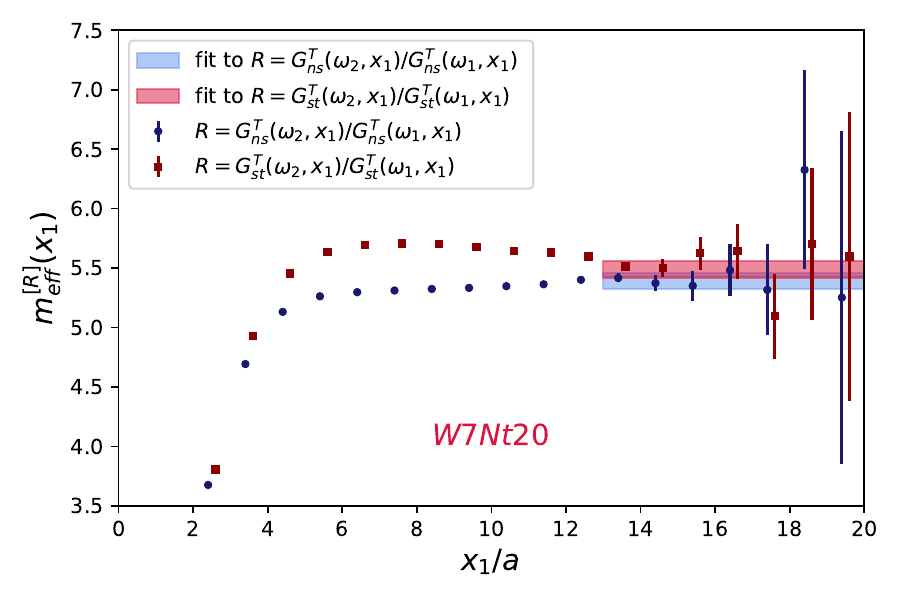}
    \caption{Effective masses (see Eq.~\eqref{eq:effmass}) for the ratios of the $n=2$ to the $n=1$ Matsubara sector
static (red bars) and non-static (blue bars) screening correlators on W7.}
    \label{fig:static_and_non_static_eff_mass_gap_W7}
\end{figure}

Since there is a systematic error associated with the choice of fit range and the number of states included, we also follow an alternative procedure where we include a prior in the fit. We obtain these priors by fitting the 
 ratios of correlators $G_i^T(\omega_2,x_1)/G_i^T(\omega_1,x_1)$
 to a single exponential. The thus extracted mass gap is shown in Fig.~\ref{fig:static_and_non_static_eff_mass_gap_W7}. When added to the screening mass associated with $G_i^T(\omega_1,x_1)$, it provides a suitable prior for the screening mass in the $\omega_2$ sector.
 The continuum-extrapolated values of these priors are
\begin{align}
    \label{eq:screening_mass_static_omega2_from_ratio}
    m_{\text{st}}^{\text{0,ratio}}(\omega_2)/T &= 14.07(18)\,,\\
    \label{eq:screening_mass_nonstatic_omega2_from_ratio}
    m_{\text{ns}}^{\text{0,ratio}}(\omega_2)/T &= 14.99(21)\,. 
\end{align}
At the finest lattice spacing, the results from this procedure are illustrated in Fig.~\ref{fig:static_and_non_static_eff_masses_X7} with the green band.

The motivation for this alternative procedure is that the local
effective masses, defined generically by 
\begin{align}
    \label{eq:effmass}
    \frac{\cosh[(x_1+a/2-L/2) m_{\text{eff}}^{[G]}(x_1)]}{\cosh[(x_1-a/2-L/2) m_{\text{eff}}^{[G]}(x_1)]}= \frac{G(x_1+a/2)}{G(x_1-a/2)}\,,
\end{align}
resemble each other in the $n=2$ and $n=1$ sector.
In the continuum and infinite-volume limit, the static screening mass is determined by the relativistic dispersion relation, 
\begin{align*}
     m_{\text{st}}^{\text{0,disp}}(\omega_2)/T = \sqrt{(m_{\text{st}}^0(\omega_0)/T)^2 + (4\pi)^2} = 13.96(2)\,,
\end{align*}
and this expectation is better fulfilled by the fits with priors. However, the difference between the two sets of results ultimately reflects the systematic uncertainty of describing the tail of the screening correlators.
We expect these two methods to offer complementary estimates that bracket the true screening mass.
We find the non-static screening mass in sector $\omega_2$, $m_{\text{ns}}^{\text{0,w/ prior}}(\omega_2)/T = 15.09(20)$ to be in good agreement with the EQCD prediction of $m_{\rm ns}^0(\omega_2)/T=15.0(2)$~\cite{Brandt:2014uda}.

Using our fitted values of $m^l_i$ and $A_i^l$, we extend the lattice data for the correlators $G_i^T(\omega_2,x_1)$ beyond the reference distance $x_1=\beta$ as a sum of exponentials, as illustrated in the top panel of Fig.~\ref{fig:H_E_omega_2_with_tails_d3_X7_and_simul_cont_extr_H_E_omega_2}, in order to compute $H_E(\omega_2)$ according to Eq.~\eqref{eq:H_E_standard_subtraction}.
We perform separate model averages (MAV) for the analyses performed without and with a prior, obtaining
\begin{align}
    \label{eq:H_E_omega2_result_with_prior_mav}
    [-H_E(\omega_2)/T^2]_{\text{w/ prior}}^{\text{MAV}} &= 0.891(24)_{\text{stat}}(54)_{\text{sys}}^\mathrm{cont}[59]\,,  \\
    \label{eq:H_E_omega2_result_wo_prior_mav}
    [-H_E(\omega_2)/T^2]_{\text{w/o prior}}^{\text{MAV}} &= 0.846(27)_{\text{stat}}(40)_{\text{sys}}^\mathrm{cont}[48]\,.
\end{align}
The results in (\ref{eq:H_E_omega2_result_with_prior_mav}-\ref{eq:H_E_omega2_result_wo_prior_mav}) agree well with those obtained from a simultaneous, uncorrelated continuum extrapolation using a common intercept across a range of $k$-values, see Fig.~\ref{fig:H_E_omega_2_with_tails_d3_X7_and_simul_cont_extr_H_E_omega_2}, bottom panel.
The systematic uncertainty is estimated via Eq.~\eqref{eq:systematic_error} and reflects the variation among continuum extrapolations obtained from different choices of $\Omega_n$; see Eq.~\eqref{eq:sine_modification} and below.

As our final result, we quote the mean of Eqs. (\ref{eq:H_E_omega2_result_with_prior_mav}-\ref{eq:H_E_omega2_result_wo_prior_mav}),
\begin{align}
    -H_E(\omega_2)/T^2 = 0.869(24)_{\text{stat}}(54)_{\text{sys}}^\mathrm{cont}(44)_{\text{sys}}^{\text{tail}}[74]_{\text{tot}}\,,
\end{align}
where the systematic uncertainty related to the tail extension is taken as the full difference of $0.044(17)$ between the two results (\ref{eq:H_E_omega2_result_with_prior_mav}-\ref{eq:H_E_omega2_result_wo_prior_mav}), which is statistically significant.

Our result is compatible with the previous result from \cite{Ce:2023oak} whilst having a smaller uncertainty.
The vastly-improved statistical precision, up to a factor 6.5 in the integrand at separations of $x_1\approx \beta$ [see Fig.~\ref{fig:comparison_integrand_O7_ws_vs_ps} in End Matter], has enabled us to make a comprehensive study of the systematic uncertainties due to the large-$x_1$ contribution.
Finally, subtracting the precisely determined first moment $-H_E(\omega_1)/T^2=0.676(7)$, we obtain 
\begin{equation}
    \label{eq:H_E_difference}
    -\left[H_E(\omega_2) - H_E(\omega_1)\right]/T^2=0.193(74)\,.
\end{equation}
The same quantity evaluated from the AMY spectral function~\cite{Arnold:2001ms} lies in the interval $[0.25,\, 0.30]$, depending on the value of the strong coupling $\alpha_\mathrm{s}\in[0.25,\,0.31]$ used. Thus, our result is on the low side of but compatible with the weak-coupling prediction.
Previous lattice studies~\cite{Ghiglieri:2016tvj,Ce:2020tmx,Ce:2022fot,Ali:2024xae} found photon emissivities in the quark-gluon plasma consistent with the AMY prediction, but we emphasize that ours is free of systematic uncertainties associated with an inverse problem, and provides evidence at the $2.6\,\sigma$ level for a non-zero photon emi\-ssi\-vity.

\textit{Conclusion.---} By computing the second moment of the photon spectrum for quark-gluon plasma at a temperature around 254\,MeV, we were able to probe the emissivity of hard photons and compare it to weak-coupling predictions. 
Our current result for $\vert H_E(\omega_2) - H_E(\omega_1)\vert$ is on the lower end of the corresponding leading-order prediction~\cite{Arnold:2001ms};
note that the next-to-leading order correction is positive~\cite{Ghiglieri:2013gia}. Furthermore,
in our earlier study~\cite{Krasniqi:2024inr}  we found that $\vert H_E(\omega_1)\vert/T^2$ near the crossover is comparable to that in the high-temperature phase, indicating sizable photon emission in this temperature regime corresponding to late times in heavy-ion collisions. These observations could shed light on the direct photon puzzle, since the azimuthal anisotropy of the photons rises if a larger fraction of the produced photons originate from the later stages of the collision~\cite{vanHees:2011vb}.
It would thus be interesting to quantitatively study the impact of our findings concerning the thermal photon emissivity on the hydrodynamics-based predictions for the photon yield and anisotropy in heavy-ion collisions at RHIC and the LHC.

We thank Csaba T\"or\"ok for a pleasant collaboration~\cite{Ce:2023oak} in the recent past, out of which this project grew.  
This work was supported by the European Research Council (ERC) under the European Union’s Horizon 2020 research and innovation program through Grant Agreement No. 771971-SIMDAMA, as well as by the Deutsche Forschungsgemeinschaft (DFG, German Research Foundation) through the Cluster of Excellence “Precision Physics, Fundamental Interactions and Structure of Matter” (PRISMA$^+$ EXC 2118/1) funded by the DFG within the German Excellence strategy (Project ID 39083149). The research of MC is funded by the Italian Ministry of University and Research (MUR) through the ``Rita Levi Montalcini'' programme for young researchers. RJH acknowledges support from the U.S. National Science Foundation under award OAC-2311430. 
The generation of gauge configurations was performed on the Clover and Himster2 platforms at Helmholtz-Institut Mainz and on Mogon II at Johannes Gutenberg University Mainz. We have also benefited from computing resources at Forschungszentrum J\"ulich allocated under NIC project HMZ21. 
The authors gratefully acknowledge the computing time provided to them on the high-performance computing cluster Noctua\;2 at the NHR Center PC2, which is funded by the Federal Ministry of Education and Research and the state governments participating on the basis of the resolutions of the GWK for the national high-performance computing at universities (www.nhr-verein.de/unsere-partner).
For generating the configurations and performing measurements, we used the openQCD~\cite{Luscher:2012av} as well as the QDP++ packages~\cite{Edwards:2004sx}, respectively.

\vspace{2mm}
\noindent$^*$\;arkrasni@uni-mainz.de
\bibliographystyle{apsrev4-2}
\bibliography{main}
\appendix

\section{End Matter}
\label{sec:lattice_parameters}
\begin{table}[b]
\caption{Overview of the $N_\mathrm{f}=2$ ensembles used in this study.
The parameters given are the bare gauge coupling $g_0$, the Wilson hopping 
parameter $\kappa$, the lattice spacing $a$, the temporal size $\beta$ in units of the lattice spacing $a$, the number of configurations used $N_{\rm con}$ and the number of inversions $N_{\text{inv}}$.}

\centering
\begin{tabular}{cccccccc}
\toprule
label & $6/g_0^2$ & $\kappa$ & $a\,$[fm] & $\beta/a$ & $N_{\rm con}$ & $\frac{N_{\text{inv}}^{\omega_1}}{10^7}$& $\frac{N_{\text{inv}}^{\omega_2}}{10^7}$  \\
\midrule
O7 & 5.5       & 0.13671        & 0.049  & 16   & 1400  & 2.15 & $8.60$ \\ 
W7 & 5.685727  & 0.136684       & 0.039  & 20   & 1600  & 1.54 & $4.61$ \\ 
X7 & 5.827160  & 0.136544       & 0.033  & 24   & 1500  & 1.73 & $5.18$\\ 
\bottomrule
\end{tabular}
\label{tab:lattice_param}
\end{table}

\begin{figure}[h]
    \centering
    \includegraphics[scale=0.53]{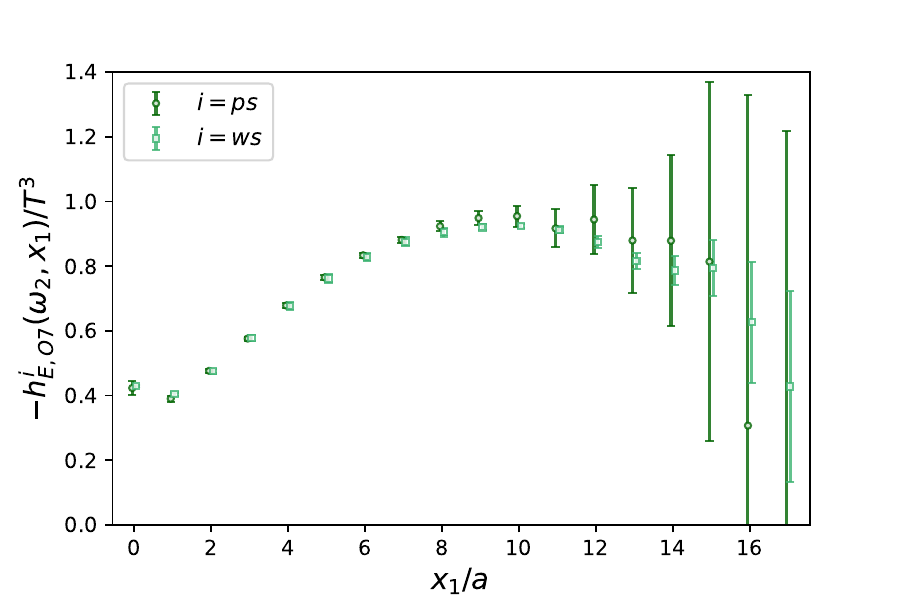}
    \caption{Comparison of the integrand $-h_{E}^{\text{ps}}(\omega_2, x_1)$ $(k=\infty)$ used in~\cite{Ce:2023oak} with the integrand $-h_{E}^{\text{ws}}(\omega_2, x_1)$ used in this work, on the coarsest ensemble O7. For instance, the error on $-h_{E}(\omega_2, x_1=15a)$ has been decreased by a factor of $\approx 6.5$.
        }
    \label{fig:comparison_integrand_O7_ws_vs_ps}
\end{figure}

\begin{figure}[]
    \centering
    \includegraphics[scale=0.53]{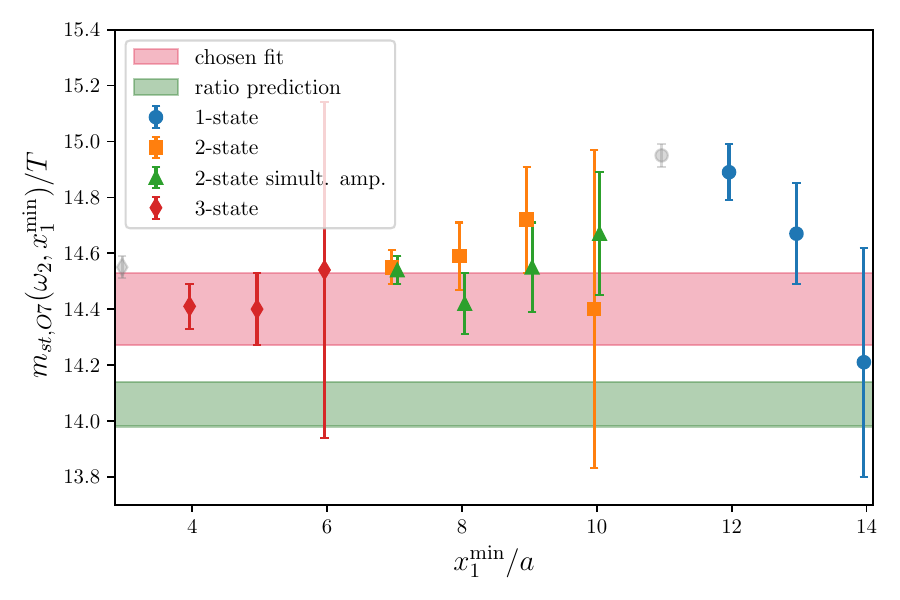}
    \includegraphics[scale=0.53]{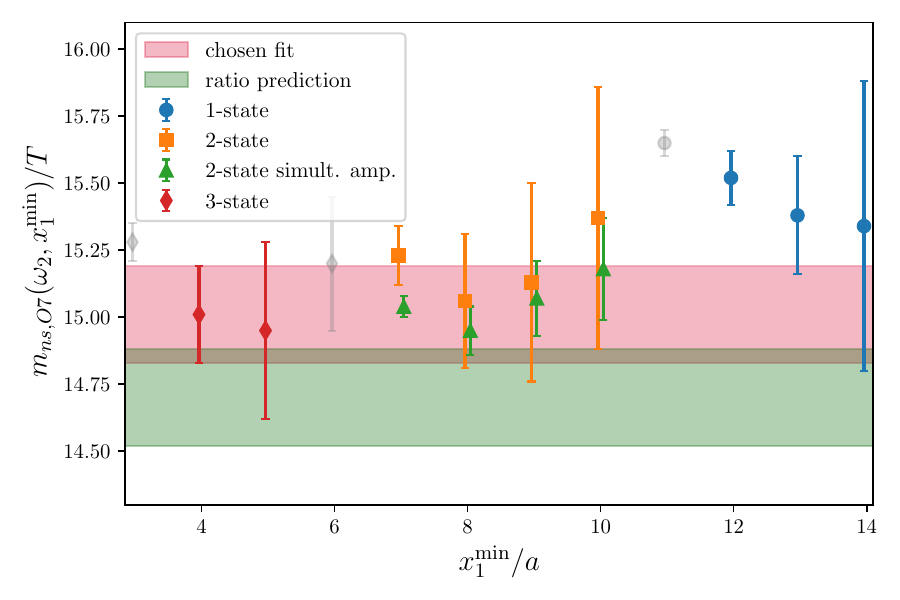}
    \caption{Fit results along with the ratio prediction [finite lattice spacing pendant of Eqs.~(\ref{eq:screening_mass_static_omega2_from_ratio}-\ref{eq:screening_mass_nonstatic_omega2_from_ratio})] for the ground-state static (top) and non-static (bottom) screening masses in the second Matsubara  channel on O7.  The fit range is set by choosing $x_1^{\text{max}}/a$ as the first point where the correlator is consistent with zero, while $x_1^{\text{min}}/a$ is varied systematically. One-, two-, and three-state fits are shown; for the two-state case, simultaneous fits to static and non-static amplitudes are also included. Grayed-out points correspond to fits with $\text{p-value}>0.95$ or $ < 0.05$.} 
    \label{fig:fit_systematic}
\end{figure}

\textit{Simulation parameters \& algorithmic aspects.---}
Simulations were performed at a fixed temperature of $T \approx 254\,\mathrm{MeV}$ and aspect ratio $L/\beta = 4$, with $L$ ($\beta$) denoting the spatial (temporal) lattice extent. We use the local discretization of the vector current, which has support on a single lattice site. We renormalize the local vector current by multiplying with $Z_V(g_0^2)$, with renormalization constants taken from from Ref.~\cite{DallaBrida:2018tpn}. The ensemble parameters are summarized in Table~\ref{tab:lattice_param}. 
The number of Dirac operator inversions is computed as
\begin{equation*}
N_{\text{inv}} = N_{\text{trans}} \times N_{\text{hits}} \times N_{\text{con}} \times N_{p} \times 4 \times 3\,,
\end{equation*}

where $N_{\text{trans}}$ denotes the number of spatial translations, (which we set to $L/a$), $N_{\text{hits}}$ are the number of stochastic draws of the source, $N_{p}$ is the number of momentum sources, the factor $4$ is the number of spin indices and  $3$ the number of spatial decay directions we use. Note that no color dilution is applied.
The impact on the integrand $h_E(\omega_2,x_1)$ of the improved statistical precision on the current-current correlators computed with wall sources (ws) instead of point sources (ps)  is illustrated in Fig.~\ref{fig:comparison_integrand_O7_ws_vs_ps}.
 \medskip

\textit{Correlator fits: varying the number of states.---}
As a consistency check, we have fitted the static and non-static correlators in sector $\omega_2$ with various numbers of states included in the ansatz Eq.\ (\ref{eq:inf_volume_fit}).
The impact of these variations on the lowest screening mass is illustrated in Fig.~\ref{fig:fit_systematic} for ensemble O7. Compatible results are found, provided the lower end $x_1^{\rm min}$ of the fit range is chosen sufficiently large.

\medskip
\textit{Lattice perturbation theory \& choice of kernel.---}

\begin{figure}[H]
    \centering
    \includegraphics[scale=0.53]{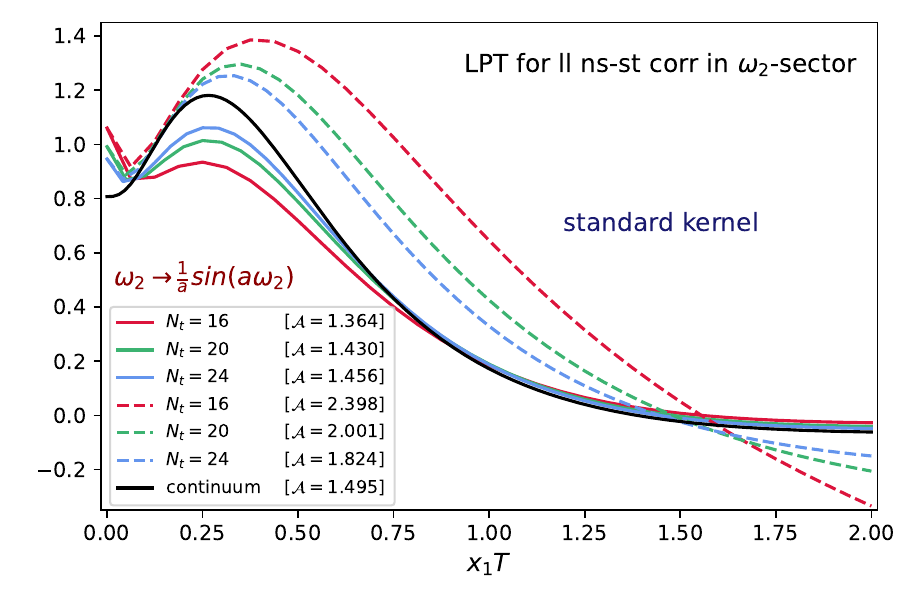}
    \includegraphics[scale=0.53]{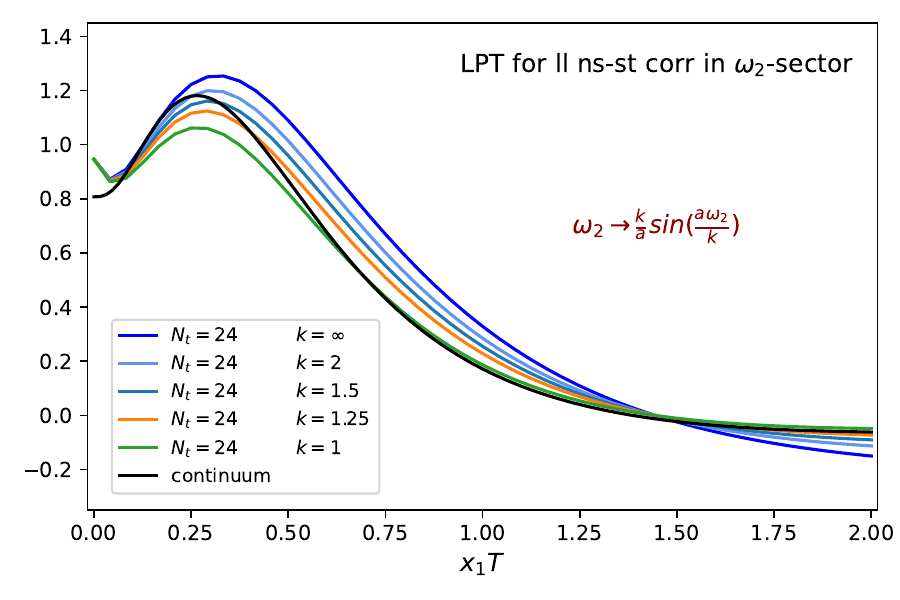}
    \caption{
        \textbf{Top panel:} Leading-order lattice perturbation theory prediction for the standard subtracted correlator (using the local discretization of the vector current) with the cosh-kernel (dashed lines) and with the modified ($k=1$) sine-kernel (solid line) in the free theory for the second Matsubara sector $\omega_2$.
        \textbf{Bottom panel:} Comparison of modifications with different $k$-values for the $\beta/a=24$ case.}
    \label{fig:cutoff_effects}
\end{figure}

In order to guide the choice of the kernel, we analyze the behavior of the integrand $h_E(\omega_2,x_1)$ in leading-order lattice perturbation theory with the local discretization of the current. Fig.~\ref{fig:cutoff_effects} (top panel) shows this quantity multiplied by the standard kernel (dashed lines) and the $k=1$ kernel (solid lines), respectively. The standard kernel leads to large cutoff effects in the intermediate region $x_1 T \in [\frac{1}{3},\frac{3}{2}]$, while the modified kernel ($k = 1$) reduces these substantially at $x_1 T > 3/4$.

Notably, with the standard kernel, the continuum limit is approached from above, whereas the modified kernel (\( k = 1 \)) approaches it from below, which suggests a potential benefit of interpolating between them. We employ the smooth Heaviside function,
\begin{equation}
    \label{eq:smooth_heaviside}
    \Theta(x, x_w, d) = \frac{1}{2} \left[ 1 + \tanh\left(\frac{x - x_w}{d}\right) \right]\,,
\end{equation}

to achieve a controlled transition between the two kernels: unmodified at short distances $x_1 < x_w$ and modified at long distances $x_1 > x_w$. For $d\approx0.15\,\mathrm{fm}$ this results in a flat continuum extrapolation, see Fig.~\ref{fig:H_E_omega_2_with_tails_d3_X7_and_simul_cont_extr_H_E_omega_2} (bottom panel).
Additionally, the bottom panel of Fig.~\ref{fig:cutoff_effects} compares different kernel modifications for the \( \beta/a = 24 \) case, corresponding to our finest ensemble, X7. It is evident that the modifications with \( k = \frac{3}{2} \) and \( k = 2 \) also reduce cutoff effects, although neither performs as well as the \( k = 1 \) prescription.

\end{document}